\documentclass[fleqn,usenatbib]{mnras}

\usepackage{newtxtext,newtxmath}

\usepackage[T1]{fontenc}

\DeclareRobustCommand{\VAN}[3]{#2}
\let\VANthebibliography\thebibliography
\def\thebibliography{\DeclareRobustCommand{\VAN}[3]{##3}\VANthebibliography}

\usepackage{graphicx}	
\usepackage{amsmath}	
\usepackage{amssymb}	

\usepackage{microtype} 

\usepackage{etoolbox, siunitx}
\robustify\bfseries

\usepackage{booktabs} 

\usepackage[normalem]{ulem} 

\newcommand{\coralie}{CORALIE}
\newcommand{\harps}{HARPS}
\newcommand{\euler}{\emph{Euler}}
\newcommand{\tess}{\emph{TESS}}

\newcommand{\pymcthree}{\textsf{pyMC3}}
\newcommand{\emcee}{\textsf{emcee}}
\newcommand{\elle}{\textsf{elle}}

\newcommand{\teff}{\ensuremath{T_\mathrm{eff}}}
\newcommand{\vsini}{\ensuremath{v_\mathrm{eq}\sin{i_{\star,1}}}}
\newcommand{\veq}{\ensuremath{v_\mathrm{eq}}}
\newcommand{\istar}{\ensuremath{i_{\star,1}}}
\newcommand{\ccfout}{CCF\ensuremath{_\mathrm{out}}}
\newcommand{\ccfin}{CCF\ensuremath{_\mathrm{in}}}

\newcommand{\Protsd}{\ensuremath{18.1 \pm 1.6\,\mathrm{d}}}

\DeclareSIUnit\au{AU}
\DeclareSIUnit\Rsun{R_\odot}
\DeclareSIUnit\Msun{M_\odot}
\DeclareSIUnit\gyr{Gyr}
\DeclareSIUnit\ppt{ppt}

\newcommand*{\ra}[2][]{{
    \def\SIUnitSymbolDegree{\textsuperscript{h}}%
    \def\SIUnitSymbolArcminute{\textsuperscript{m}}%
    \def\SIUnitSymbolArcsecond{\textsuperscript{s}}%
    \ang[#1]{#2}
    }%
}

\title[Spin--orbit alignment for EBLM J0608-59]{The EBLM project. VII.
Spin--orbit alignment for the circumbinary planet host \mbox{EBLM
J0608-59\,A/TOI-1338\,A}}

\author[V. Kunovac Hod\v{z}i\'c et al.]{%
        Vedad Kunovac Hod\v{z}i\'c$^{1,2}$\thanks{\href{mailto:vxh710@bham.ac.uk}{vxh710@bham.ac.uk}}\thanks{Fulbright Fellow},
        Amaury H.\,M.\,J. Triaud$^{1}$,
        David V. Martin$^{3}$\thanks{Fellow of the Swiss National Science Foundation},\newauthor
        Daniel C. Fabrycky$^{2}$,
        Heather M. Cegla$^{4,5}$,
        Andrew Collier Cameron$^{6}$,
        Samuel Gill$^{4}$,\newauthor
        Coel Hellier$^{7}$,
        Veselin B. Kostov$^{8,9}$, 
        Pierre F.\,L. Maxted$^{7}$,
        Jerome A. Orosz$^{10}$,\newauthor
        Francesco Pepe$^{5}$,
        Don Pollacco$^{4}$, 
        Didier Queloz$^{11,5}$,
        Damien S\'egransan$^{5}$,\newauthor
        St\'ephane Udry$^{5}$ 
        and William F. Welsh$^{10}$
\\
\hfill \\
$^{1}$School of Physics and Astronomy, University of Birmingham, Edgbaston, Birmingham B15 2TT, UK\\
$^{2}$Department of Astronomy and Astrophysics, University of Chicago, 5640 S. Ellis Avenue, Chicago, IL 60637, USA\\
$^{3}$Department of Astronomy, The Ohio State University, 4055 McPherson Laboratory, Columbus, OH 43210, USA\\
$^{4}$Department of Physics, University of Warwick, Gibbet Hill Road, Coventry CV4 7AL, United Kingdom\\
$^{5}$Observatoire Astronomique de l’Universit\'e de Gen\`eve, 51 Chemin des Maillettes, CH-1290 Sauverny, Switzerland\\
$^{6}$SUPA, School of Physics \& Astronomy, University of St Andrews, North Haugh, KY16 9SS, St Andrews, Fife, Scotland, UK\\
$^{7}$Astrophysics Group, Keele University, Staffordshire, ST5 5BG, UK\\
$^{8}$NASA Goddard Space Flight Center, 8800 Greenbelt Road, Greenbelt, MD 20771, USA\\
$^{9}$SETI Institute, 189 Bernardo Avenue, Suite 200, Mountain View, CA 94043, USA\\
$^{10}$Department of Astronomy, San Diego State University, 5500 Campanile Drive, San Diego, CA 92182, USA\\
$^{11}$Cavendish Laboratory, J J Thomson Avenue, Cambridge, CB3 0HE, UK
}

\date{Accepted XXX. Received YYY; in original form ZZZ}

\pubyear{2015}

\begin{document}

\label{firstpage}
\pagerange{\pageref{firstpage}--\pageref{lastpage}}
\maketitle

\begin{abstract} A dozen short-period detached binaries are known to host
    transiting circumbinary planets. In all circumbinary systems so far, the
    planetary and binary orbits are aligned within a couple of degrees. However,
    the obliquity of the primary star, which is an important tracer of their
    formation, evolution, and tidal history, has only been measured in one
    circumbinary system until now.  EBLM J0608-59/TOI-1338 is a low-mass
    eclipsing binary system with a recently discovered circumbinary planet
    identified by \tess{}. Here, we perform high-resolution spectroscopy during
    primary eclipse to measure the projected stellar obliquity of the primary
    component.  The obliquity is low, and thus the primary star is aligned with
    the binary and planetary orbits with a projected spin--orbit angle $\beta =
    \ang[angle-symbol-over-decimal]{2.8} \pm
    \ang[angle-symbol-over-decimal]{17.1}$.  The rotation period of
    \SI[separate-uncertainty,multi-part-units=single]{18.1 \pm 1.6}{\day}
    implied by our measurement of $v\sin{i_\star}$ suggests that the primary has
    not yet pseudo-synchronized with the binary orbit, but is consistent with
    gyrochronology and weak tidal interaction with the binary companion.  Our
    result, combined with the known coplanarity of the binary and planet orbits,
    is suggestive of formation from a single disc. Finally,
    we considered whether the spectrum of the faint secondary star could affect
    our measurements.  We show through simulations that the effect is negligible
    for our system, but can lead to strong biases in $v\sin{i_\star}$ and
    $\beta$ for higher flux ratios. We encourage future studies in eclipse
    spectroscopy test the assumption of a dark secondary for flux ratios
    $\gtrsim$\SI{1}{\ppt}.
\end{abstract}




\section{Introduction}\label{sec:introduction}

\begin{table}

    \centering
    \caption{Target and binary orbital parameters. Subscripts 1 and 2 denote the primary and secondary components, respectively. Brackets denote uncertainties on last two digits. $^{a)}$Determined from rotational line broadening using \harps{} \citep{kostov2020}. $^{b)}x = \num{2450000}$.} 
    \footnotesize
    \begin{tabular*}{\columnwidth}{@{\extracolsep{\fill}}
        l
        l
        l
        }
        \toprule
        \toprule
        Parameter & Description & Value \\
        \midrule
        \multicolumn{3}{@{}l@{}}{\emph{Target information}} \\
        $\alpha$ & Right ascension & \ra[angle-symbol-over-decimal,minimum-integer-digits=2]{06;08;31.95} \\
        $\delta$ & Declination & \ang[angle-symbol-over-decimal]{-59;32;28.1} \\
        $V_\mathrm{mag}$ & Apparent magnitude & 11.73 \\[5pt]
        
        \multicolumn{3}{@{}l@{}}{\emph{Stellar parameters}} \\
        $M_1$ (\si{\Msun}) & Primary mass & 1.127(68) \\
        $M_2$ (\si{\Msun}) & Secondary mass & 0.313(11) \\
        $R_1$ (\si{\Rsun}) & Primary radius & 1.331(24) \\
        $R_2$ (\si{\Rsun}) & Secondary radius & 0.3089(56) \\
        $T_{\rm eff,1}$ (\si{\kelvin}) & Primary temperature & 6050(80) \\
        $T_{\rm eff,2}$ (\si{\kelvin}) & Secondary temperature & 3330(50) \\
        $\log{g_1}$ (cgs) & Surface gravity & 4.00(08) \\
        $[\mathrm{Fe}/\mathrm{H}]_1$ & Metallicity & 0.01(05) \\
        $^{a}$\vsini{} (\si{\kilo\metre\per\second}) & Projected rotation & 3.6(0.6) \\
        Age (Gyr) & Isochrone binary age & 4.4(0.2) \\[5pt]

        \multicolumn{3}{@{}l@{}}{\emph{Orbital parameters}} \\
        $P_{\rm bin}$ (\si{\day}) & Binary orbital period & 14.608559(13) \\
        $T_\mathrm{pri}$ (BJD$-x^{b}$) & Time of primary eclipse & 53336.8242(25) \\
        $K_1$ (\si{\kilo\metre\per\second}) & RV semi-amplitude & 21.6247(34) \\
        $e_{\rm bin}$ & Eccentricity & 0.15603(15) \\
        $\omega_{\rm bin}$ (\si{\degree}) & Periastron angle & 117.554(72) \\
        $i_{\rm bin}$ (\si{\degree}) & Orbital inclination & 89.70(18) \\
        $b_\mathrm{bin}$ (R$_1$) & Impact parameter & 0.097(57) \\
        $a_\mathrm{bin}$ (\si{\au}) & Separation & 0.1321(24) \\
        $J_\tess$ & Surface brightness ratio & 0.0926(28) \\
        $u$ & Lin. limb darkening coeff. & 0.40 \\
        $v$ & Quad. limb darkening coeff. & 0.29 \\
        \bottomrule
    \end{tabular*}
    \label{tab:inputparams}
\end{table}

EBLM J0608-59 consists of an inner low-mass binary ($M_1 = \SI{1.13}{\Msun}$,
$M_2 = \SI{0.31}{\Msun}$) on a 14.6-day  eccentric orbit
\mbox{\citep{triaud2017}},
with an outer Saturn-sized circumbinary planet on a 95-day orbit recently
discovered in \tess{} photometry (TOI-1338\,AB\,b, \citealt{kostov2020}). 
hereafter Kostov et al. (2020, under rev.)).  The inner binary was first
identified as a transiting planet candidate by the WASP survey
\citep{pollacco2006}, but radial-velocities measurements soon determined it was
instead a G+M single-lined eclipsing binary. Subsequently, EBLM J0608-59
(hereafter J0608-59) was followed up spectroscopically as part of the EBLM
project whose goal is to study the properties of low-mass eclipsing binary
systems \citep{triaud2013,triaud2017}. One of the objectives of the EBLM project
is to provide observational constraints on the tidal physics of tight binaries.
To study this, we collect three types of observables:  rotational velocities
\vsini{} of the primary stars, precise eccentricities and, in some cases,
spin--orbit angles $\beta$\footnote{the notation $\lambda$ has been used for the
spin--orbit angle of exoplanets, with $\lambda = - \beta $. $\beta$, defined by
\citet{kopal1942}, i.e. the projected angle between the orbital and stellar spin
axes, is widely used within the binary star community.}, which are obtained by
measuring the Rossiter-McLaughlin effect
(\citealt{rossiter1924,mclaughlin1924,kopal1942}).

Despite many spin--orbit measurements obtained on transiting exoplanets (e.g.
\citealt{winn2015,triaud2018}), not many have been obtained on eclipsing
binaries.
A list of historical measurements can be found in
\citet{albrecht2011}. \citet{hale1994} notes that binaries with separation
\SIrange{>30}{40}{\au} are usually found with random spins. For tighter
binaries, the most extensive collection of measurements were produced by the
BANANA  survey (e.g. \citealt{albrecht2014}), which mainly targets massive
stars, with a recent compilation from the Torun project \citep{sybilski2018}. So
far, most binary pairs appear spin--orbit aligned except DI Hercules
\citep{albrecht2009}, CV Velorum \citep{albrecht2014}, and AI Phe
\citep{sybilski2018}. The binary sample covered by the EBLM project is distinct
from these other efforts in focusing exclusively on small ($M_2/M_1 \lesssim
0.3$) mass ratio binaries with Solar-like primaries. So far three measurements
of $\beta$ have been published by the project: WASP-30 and EBLM J1219-39
\citep{triaud2013}; and EBLM J0218-31 \citep{gill2019}, where all primary stars
were found to be coplanar with their respective orbits. We have collected
Rossiter-McLaughlin measurements on dozens of systems whose analysis are
ongoing, exploring parameter space in terms of \vsini{}, eccentricities and
orbital periods. 

In this paper, we report a spectroscopic primary eclipse of J0608-59 to measure
the stellar obliquity of the primary component with respect to the binary and
planet orbits from the Rossiter-McLaughlin effect. J0608-59 is only the second
primary star host to a circumbinary planet to have a Rossiter-McLaughlin
measurement made, following the 41 day binary Kepler-16 which was found to be
aligned \citep{winn2011}.  Additionally, the \SI{7.5}{\day} binary Kepler-47 is
also thought to be spin--orbit aligned, based on eclipse spot--crossings
\citep{orosz2012}.

\section{Observations}\label{sec:observations}

A single spectroscopic primary eclipse of J0608-59 was observed on 2010 November
2 using the $1.2$-m Swiss \euler{} Telescope at La Silla, Chile. The \coralie{}
instrument spans the visible range (\SIrange{390}{680}{\nano\metre}) with an
average resolving power of $R{\sim}\num{55000}$. We obtained 19 exposures of
\SI{900}{\second} each\footnote{Except the two initial points, which were
\SI{600}{\second}.} over \SI{5.8}{\hour}, capturing the primary eclipse
entirely, including two spectra before ingress. The data were reduced using the
standard \coralie{} Data Reduction Software (DRS, \citealt{lovis2007}). In
brief, a cross correlation function (CCF) is derived between the observed
spectra and a G2 numerical mask. A comparison is made with a reference
thorium-argon spectrum, which allows for corrections based on instrumental
variations throughout the eclipse (e.g. atmospheric variations since \coralie{}
is not pressure-stabilized). Overall, \coralie{} is stable to
\SI{5}{\metre\per\second}, while the median radial velocity error per
observation is \SI{26}{\metre\per\second}.

As a part of the EBLM program, J0608-59 received an additional 19 radial
velocity measurements to map out its orbit with \coralie{}. From these data, we
select three epochs from 2009 November 28--29 and 2010 January 6 as reference
spectra due to insufficient data outside the spectroscopic eclipse night.
J0608-59 was also selected as a target in the BEBOP radial velocity search for
circumbinary planets. An additional 17 \coralie{} measurements were made and
published in \citet{martin2019}, and an additional 7 \harps{} measurements have
been taken since. Finally, J0608-59 was observed in 12 \tess{} sectors
\citep{kostov2020}, 9 of which were short-cadence data under the Guest
Investigator Program G011278 (PI: O. Turner). The BEBOP data were not used in
the analysis of this work, but we use orbital parameters
(Table~\ref{tab:inputparams}) from \citet{kostov2020} that are based on the
BEBOP and \tess{} data.

\section{Line profile analysis}\label{sec:analysis}
Radial velocity measurements are typically determined from a cross-correlation
technique. The resulting CCFs are fitted with Gaussian profiles to determine the
effective radial velocity, i.e. treating any distortion of the line profiles as
a pure Doppler shift of the line centre. The Rossiter-McLaughlin effect is
however a spectroscopic effect where the line profiles are distorted depending
on the position of the occulting body on the stellar disc.

While many techniques have been put forth to improve on the modelling of the
anomalous radial velocity during a spectroscopic transit/eclipse to account for
these effects (e.g. \citealt{albrecht2007,hirano2011a}),  other methods have
been developed for modelling the spectral line distortions directly. These
methods are commonly referred to as Doppler tomography or Doppler shadow
\citep{colliercameron2010,cegla2016,hirano2020}. Here we use the
\textit{reloaded} Rossiter-McLaughlin technique from \citet{cegla2016} to
directly retrieve the occulted light from the stellar disc with no assumptions
on the shape of the line profiles compared to other methods. For this and future
analyses we have written a dedicated software package,
\elle{}\footnote{\url{https://github.com/vedad/elle}}, and plan to release a
user-friendly public version in a forthcoming paper.
\subsection{Retrieving the occulted light} \label{sec:methods_occulted_light}
We work on the disc-integrated CCFs output from the \coralie{} DRS.  We start by
removing the Keplerian velocity due to the binary orbit by resampling the CCFs
on a common velocity grid at the spectrograph resolution of \SI{\sim
1.8}{\kilo\metre\per\second}. 
The orbital parameters to compute the Keplerian model are obtained from
\citet{kostov2020}, who  analysed the full \tess{} Cycle 1 photometry and
\coralie{}/\harps{} radial velocities, modelling both the binary orbit as well
as the circumbinary planet orbit using a photodynamical model. Their fit did not
detect any significant binary apsidal motion, and produced a $\chi_\nu^2 \approx
1$, indicating very good agreement on the data despite that ${\sim}\num{10}$
years separating the first \coralie{} observations from the \tess{} photometry.
From this we can rule out any significant changes in the binary ephemeris, such
as radial velocity drifts and associated light travel time effects due to
distant massive companions.

The CCFs are scaled by their respective continuum flux to compare their relative
flux variations. We define the continuum level as the median flux $4\sigma$
outside the computed line profile centres (radial velocity), where $\sigma$
denotes the width of the Gaussian fit from the \coralie{} DRS.  In the same
fashion we also estimate the CCF error by computing the standard deviation of
the continuum flux.  Furthermore, we also normalise the CCFs by the theoretical
primary eclipse limb-darkened light curve to account for the loss of light
during eclipse, where the light curve was again computed using orbital elements
from \citet{kostov2020}.

As with any Doppler shadow method, the in-eclipse CCFs need to be compared to a
high SNR template CCF (typically an average over several out-of-eclipse spectra)
that represents the intrinsic rotationally broadened stellar spectrum.  The
Rossiter-McLaughlin sequence we use in this work only has two spectra taken
outside of the eclipse window, both observed at high airmass, $z > \num{1.8}$,
and at shorter integration times than in-eclipse observations. Upon inspection
of the normalised CCFs we noticed -- other than lower SNR -- that these two
observations had CCF contrasts (depths) \SIrange{3}{4}{\percent} smaller than
the remaining sample from the same night.

\begin{figure}
    \centering
    \includegraphics[width=\linewidth]{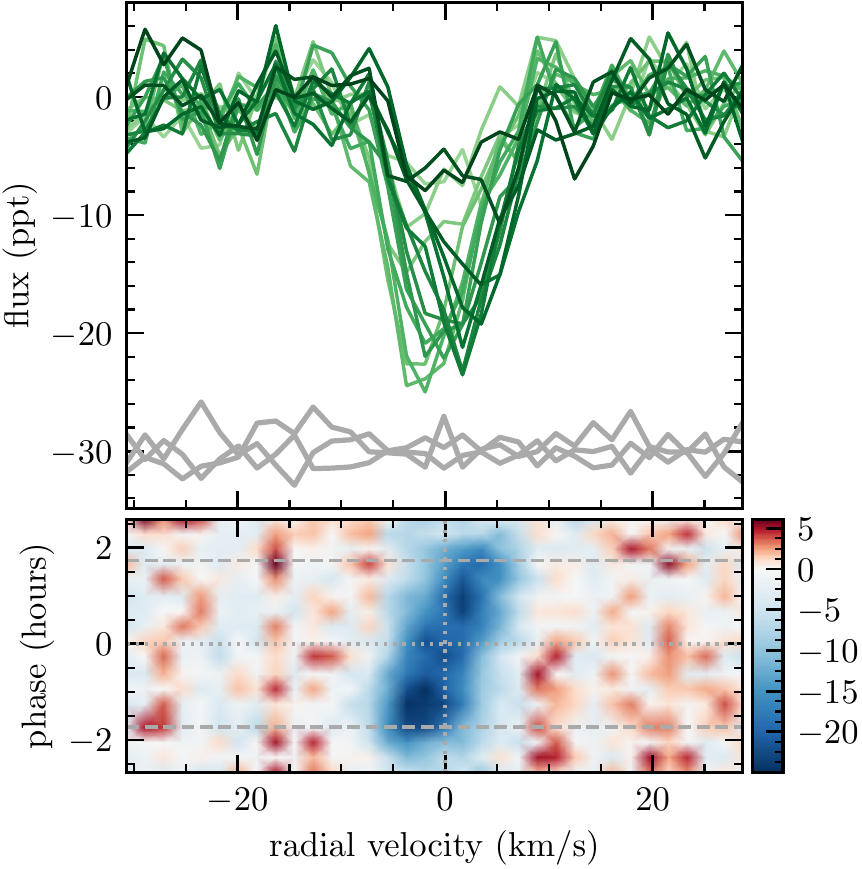}
    \caption{\emph{Upper panel:} The retrieved residual line profiles hidden
    beneath the eclipsing secondary star, with colours getting darker with time.
The grey profiles at the bottom are the residuals from the reference spectra
with an arbitrary offset. \emph{Lower panel:} Another view showing the residual
flux as a function of time on the vertical axis, displaying the trace, or
Doppler shadow, of the eclipsing body moving across the rotating stellar disc.
The grey dashed lines denote the second and third contact points, and the dotted
lines denote the mid-transit point and minimum limb angle. The colourbar shows
the CCF flux.}
\label{fig:trace}
\end{figure}

To avoid biasing our measurements due to the anomalous CCF contrast, we do not
use the two spectra at the beginning of night to build our reference. Instead,
we average the two spectra together to determine the systemic velocity at the
night of the spectroscopic eclipse, $\gamma_\mathrm{RM}$. Then, we identify
three individual spectra taken at different nights with the same exposure time
as our in-eclipse observations, and verify that their CCF contrast, FWHM, and
SNR match the data taken during eclipse. The three reference spectra are
resampled to the same grid as with the in-eclipse spectra, but shifted to
$\gamma_\mathrm{RM}$ to account for any stellar activity between the
observations. The three spectra are then averaged together, weighted by their
errors, to create a master out-of-eclipse template, \ccfout{}. The in-eclipse
spectra and \ccfout{} -- now in the same reference frame -- are then shifted by
$\gamma_\mathrm{RM}$ to bring the CCFs to the stellar rest frame. The residual
line profiles are then calculated from the difference between \ccfout{} and the
in-eclipse, disc-integrated CCFs. The resulting in-eclipse residual
profiles\footnote{These are no longer disc-integrated profiles since the
disc-integrated contribution has been removed.}, hereafter \ccfin{}, represent
the light on the stellar disc occulted by the star and are shown in
Fig.~\ref{fig:trace}, clearly displaying the trace of the secondary star as it
moves across the primary's disc.

We compute their local radial velocity by fitting Gaussian profiles to the
\ccfin{}. In order to obtain realistic uncertainties and avoid fitting spurious
signals (particularly an issue for low SNR data, such as profiles at the stellar
limb) we use a Markov chain Monte Carlo (MCMC) sampling method to explore the
full posterior distribution and propagate uncertainties in the
``nuisance''\footnote{While the depths and widths of the Gaussian can be useful
for diagnostic purposes, we are mainly interested in the line profile centre,
$\mu$, and its uncertainty.} parameters to the final radial velocity. We build
our Gaussian model in the \pymcthree{} framework \citep{salvatier2016}, varying
the line centre, $\mu$; width, $\sigma$; contrast $A$; continuum level, $c$, and
CCF error, $\epsilon$. We sample these five parameters using the No-U-Turn
(NUTS) sampler \citep{hoffman2014}. In order to avoid cherry-picking the local
profiles that provide a good fit (and thus keep for further analysis), we use
wide, informative priors on our parameters aimed at returning a conservative
estimate of the radial velocity in case of a non-detection of the local line
centre. The priors for $A$, $\sigma$, and $\epsilon$ are drawn from the
half-normal distribution as they are both restricted to positive values, and
draw from the normal distribution for $\mu$ and $c$. 

\begin{figure*}
    \centering
    \includegraphics{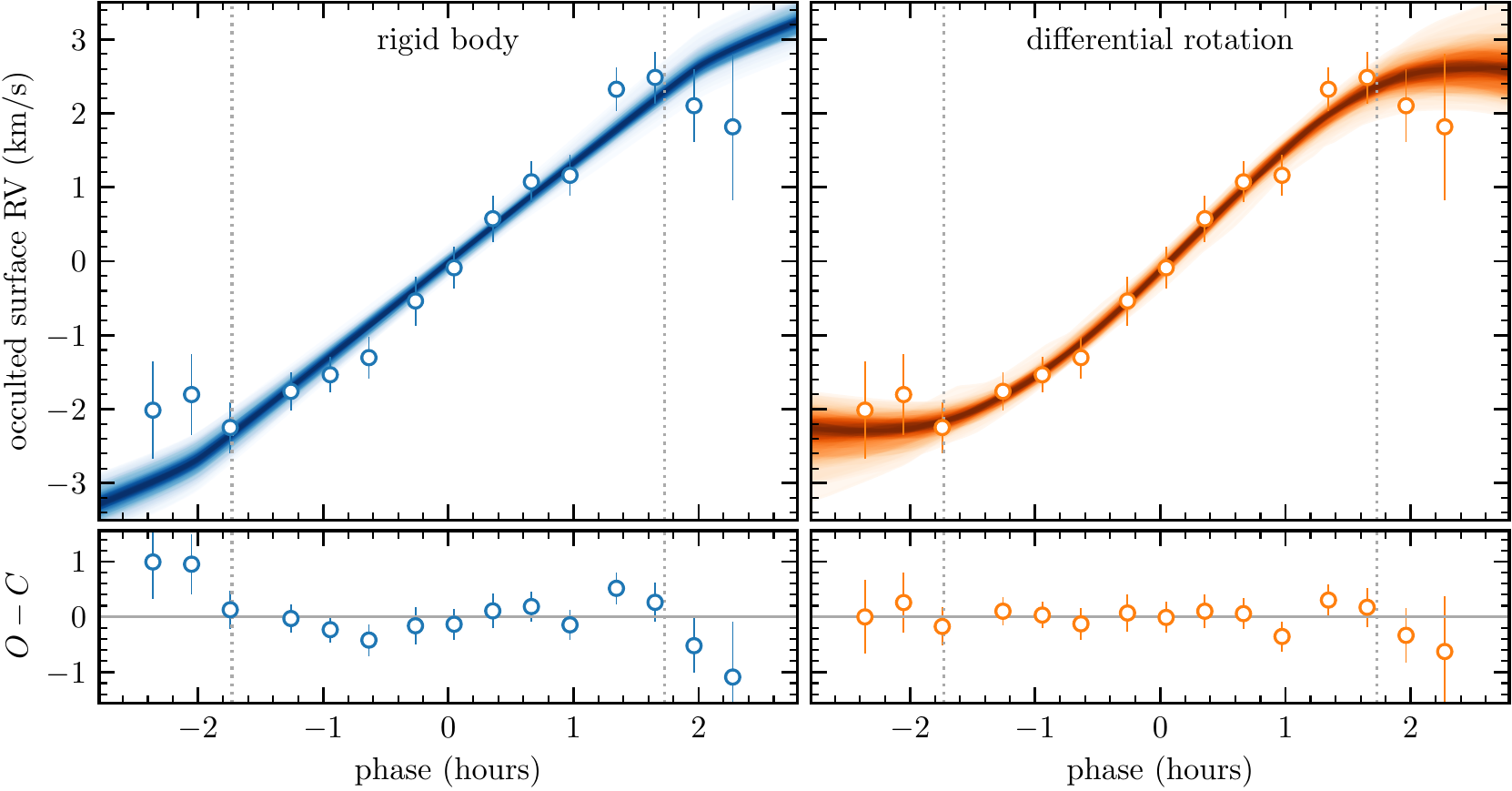}
    \caption{\emph{Upper panels:} The local radial velocities during eclipse,
        obtained from Gaussian fits to the residual profiles in
        Fig.~\ref{fig:trace}, for the rigid body model (\emph{left}) and
        differential rotation model (\emph{right}). The shading denotes the 50th
        -- 99th percentiles of the models. Dotted lines denote the second and
        third contacts, while the horizontal axis is delineated by the eclipse
        duration. \emph{Lower panels:} Residuals from the maximum likelihood
    fits of each model. The differential rotation model reduces $\chi_\nu^2$
from $1.2$ to $0.5$, and improves the Bayesian Information Criterion (BIC) by
5.5.}
    \label{fig:fit}
\end{figure*}
\subsection{Surface velocity modelling}
The surface velocity model is computed using the semi-analytical prescription in
\citet{cegla2016}. Here, we create a $51 \times 51$ grid that spans the size of
the eclipsing body, and co-moving with its centre.  At every epoch the
brightness-weighted rotational velocity at a given position is computed by
summing the cells covering the stellar disc occulted by the eclipsing body. We
assume the star follows a quadratic limb darkening function, with coefficients
$c_1 = 0.40$, $c_2 = 0.29$ obtained from interpolating \teff{}, $\log{g}$, and
$[\mathrm{Fe}/\mathrm{H}]$ in the $V$ band tables from \citet{claret2011}, using
the online tool from
\citep{eastman2013}\footnote{\url{http://astroutils.astronomy.ohio-state.edu/exofast/limbdark.shtml}}.
Oversampling and averaging our model within each \SI{900}{\second} exposure did
not impact our results.

The theoretical surface velocity at any point on a rotating rigid body is
computed from a combination of the orbital phase, $\phi(P, T_0)$; orbital
inclination, $i_\mathrm{orb}$; the stellar (primary) radius scaled by the binary
separation, $R_1/a$; projected stellar equatorial velocity, \vsini{};  and the
projected spin--orbit angle, $\beta$. While many of these parameters are known
very precisely from \citet{kostov2020} (see Table~\ref{tab:inputparams}), the
impact parameter $b = a \cos{i_\mathrm{bin}}/R_1$ is known to correlate with
both \vsini{} and $\beta$ when $b$ is close to zero. We therefore vary $R_1/a$
and $i_\mathrm{orb}$ with their Gaussian uncertainties during the fit, and let
\vsini{} and $\beta$ float freely. We fix the ephemerides $P$, $T_0$, but
checked that our results were insensitive to their influence by perturbing them
by their $1\sigma$ uncertainties.  We sample these four parameters using the
\emcee{} MCMC sampler \citep{foreman-mackey2013} to obtain their full posterior
distributions, using 200 walkers that are run for $\num{\sim 50}$ times the
autocorrelation length. The final posterior distributions are obtained after
discarding a number of burn-in steps determined visually, and thinning the
individual walkers by the autocorrelation length of the parameters. We also
attempt to model the data assuming the rotation rate varies as a function of
stellar latitude (differential rotation). In this case we can independently
sample the equatorial velocity, \veq{}, and the stellar inclination, \istar{}.
Additionally, we sample the relative shear, $\alpha = (\Omega_\mathrm{eq} -
\Omega_\mathrm{pole}) / \Omega_\mathrm{eq}$, which describes the relative
rotation rate between the poles and the equator\footnote{$\alpha \SI{\sim
0.2}{}$ for the Sun.}. We varied the  parameters uniformly in $\veq \in [0,
5.3]\,\si{\kilo\metre\per\second}$, $\beta \in [-180, 180]\,\si{\degree}$,
$\istar \in [0, 180]\,\si{\degree}$, and $\alpha \in [-2, 1]$, where $\alpha <
0$ signifies anti-solar differential rotation, i.e. that the polar latitudes
rotate faster than the equator. The upper limit on \veq{} was set to the
predicted pseudo-synchronous rotation period, which is expected at
\SI{12.7}{\day}.

\sisetup{detect-weight=true,detect-inline-weight=math}
\begin{table*}
    \renewcommand{\arraystretch}{1.2}
    \footnotesize
    \centering
    \caption{Derived spin-orbit parameters. Bold values denote adopted values and model. $^a$Fixed under rigid body assumption. $^b$\SI{68}{\percent} confidence interval.} 
    \begin{tabular*}{\linewidth}{@{\extracolsep{\fill}}
        l
        c
        c
        c
        c
        c
        c
        S
        S
        }
        \toprule
        \toprule
        Model & {\veq{} (\si{\kilo\metre\per\second})} & {\istar{} (\si{\degree})} & {$\beta$ (\si{\degree})} & {$\alpha$} & {$\psi$ (\si{\degree})} & {$P_\mathrm{rot}$ (d)} &
        {$\chi^2$} & {BIC} \\
        \midrule
        \textit{\textbf{Rigid body}} &  $\mathbf{3.69^{+0.35}_{-0.25}}$ & $\mathbf{90}^a$ & $\mathbf{2.9^{+16.0}_{-15.8}}$ & $\mathbf{0}^a$ & {--} & $\mathbf{18.3^{+1.4}_{-1.6}}$ & \bfseries 15.9 & \bfseries -9.2 \\
        \textit{Differential rotation} & $3.89^{+0.87}_{-0.68}$ & $137.1^{+13.7}_{-74.8}$  & $9.7^{+12.4}_{-11.8}$ & $[-1.85, -0.91]^b$ & $44.7^{+11.0}_{-10.5}$ & $17.3^{+3.7}_{-3.2}$ & 4.9 & -14.8 \\
        \bottomrule
    \end{tabular*}
    \label{tab:result_obliquity}
\end{table*}

\section{Results and Discussion}\label{sec:discussion}

\sisetup{separate-uncertainty, multi-part-units=single}
\subsection{Stellar obliquity and rotation}

The trace of the secondary star in Fig.~\ref{fig:trace}
    is clearly indicative of a prograde orbit as it moves from blue-shifted to
    red-shifted areas on the stellar disc of the primary. We summarise our
    results in Table~\ref{tab:result_obliquity}.  From the data modelling
    outlined in Section~\ref{sec:analysis} we find a projected rotation 
and projected stellar obliquity, $\beta = \ang[angle-symbol-over-decimal]{2.8}
\pm \ang[angle-symbol-over-decimal]{17.1}$ when assuming rigid body rotation.
The three largest sources of angular momentum in the system are the binary and
planetary orbits, followed by the primary's rotation -- whose magnitudes
contribute in approximate ratios of 6000:4:1. Given that the binary and planet
orbits are co-planar, with mutual a inclination $|\Delta i| =
\SI{0.3}{\degree}$, the largest angular momentum vectors are thus found to be
aligned. \citet{kostov2020} reported $\vsini{} = \SI{3.6 +-
0.6}{\kilo\metre\per\second}$ from rotational broadening, based on
high-resolution spectra from \harps{} obtained through the BEBOP survey (ESO
prog. ID 1101.C-0721, PI: Triaud; \citealt{martin2019}), thus the two
measurements agree very well. Our measurement of \vsini{} combined with the
stellar radius suggests a present-day rotation period of \SI{18.1 \pm
1.6}{\day}, assuming a stellar inclination of \SI{90}{\degree}.

The rigid body model in the left panel of Fig.~\ref{fig:fit} does not accurately
predict the surface velocity at the limbs. One could rightly expect the
behaviour at the limb to depend on the particular choice of limb darkening
models and its parameters, as well as the size of planet grid. We addressed each
of these scenarios by varying the limb darkening coefficients within Gaussian
uncertainties of 0.1, using a power-2 limb darkening law (e.g.
\citealt{maxted2018}), and increasing the planet grid size up to $91 \times 91$.
None made any detectable impact on our results. Moreover, the centre-to-limb
variation due to convective blueshift \citep{cegla2016,cegla2016a} is expected
to be symmetric around the mid-transit time, while the data at both limbs seem
to be anti-symmetric (dragged towards zero velocity), and thus can not explain
the effect. Most likely the effect is due to correlated noise of an unknown
origin, possibly originating from nightly differences between the reference and
eclipse spectra. Nevertheless, we verified that our results stayed consistent
when only fitting the data between the 2nd and 3rd eclipse contact points,
obtaining differences within the $1\sigma$ uncertainties of the parameters.

\subsection{Formation scenario}

Tidal evolution has three relevant effects on a binary: $i$) alignment of
stellar spin axes; $ii$) synchronisation of stellar rotation rates; and $iii$)
circularization of the orbit. The first two effects are believed to occur on a
roughly similar timescale, whereas the circularization is a much slower process,
owing to there being significantly more angular momentum in the orbit than in
the stars \citep{hut1981}. J0608-59 is too widely separated ($R_1/a_{\rm
bin}=0.023$) for circularization to have occurred within our estimated lifetime
of \SI{\sim 4}{\gyr}\footnote{Derived in \citet{kostov2020} using stellar
isochrones.}, and so its eccentricity of $e_{\rm bin}=0.156$ is not surprising. 

For non-circular binaries, such as J0608-59, tides do not synchronize the
rotation rates with the orbital period (\SI{14.61}{\day}). Instead, the stellar
rotation will over time pseudo-synchronise with the orbital motion at periastron
\citep{hut1981}. Using their Eq. (42), the pseudo-synchronous rotation period is
\SI{12.7}{\day} for J0608-59. This is significantly shorter than our derived
rotation period of
\Protsd{}.  On the other hand, using the gyrochronological age-rotation relation
for single stars from \citet{mamajek2008}, the predicted rotation period from
stellar spin-down is estimated to be \SI{19.6 +- 2.4}{\day} for our \SI{\sim
4}{\gyr} star, which may provide a plausible  explanation for the measured
rotation rate given the binary separation. We therefore suggest that the primary
star's rotation rate is largely unaffected by tides from the binary companion,
but rather driven by magnetic braking. This result is consistent with Figure 10
of \citet{torres2010}, showing that binaries with $R_1/a \lesssim 0.1$ are not
necessarily pseudo-synchronised, and also remain eccentric.

The fact that the star is not pseudo-synchronised may appear at odds with an
aligned projected obliquity of \SI{\sim 0}{\degree}. However, we present two
simple explanations for this apparent discrepancy. First, the stellar obliquity
may too be unaffected by tides, but rather the binary was primordially aligned
and such alignment persisted through its evolution. This points to the two stars
having formed from gravitational fragmentation within a single disc. If they
formed at an initially wider separation (predicted by e.g. \citealt{bate2002}),
then their orbital shrinkage would be due to accretion and disc migration, and
not a more violent scattering event, such that the stellar alignment and
circumbinary planet are preserved \citep{martin2019a}. An alternate explanation
is that the timescale of spin alignment, at least in the case of J0608-59, is
noticeably shorter than that of pseudo-synchronisation.

The Kepler-16 circumbinary planet system, by comparison, has a wider binary
($R_1/a_{\rm bin}=0.013$) that is also eccentric and spin-orbit aligned, but
contrarily has a rotation period equal to that expected from
pseudo-synchronisation (\SI{\sim 35}{\day}). However, \citet{winn2011} point out
that this rotation rate also agrees well with the expectation from
gyrochronology. Our result may support the view that the rotation of
Kepler-16\,A is only coincidental with the pseudo-synchronous rotation period
and has not been synchronised by tides, but has rather spun down to its present
rate due to the natural spin-down of stars from magnetic braking. The Kepler-47
circumbinary system is comparatively much tighter ($R_1/a_{\rm bin}=0.053$) than
both J0608-59 and Kepler-16. Tides are most likely responsible for its small
eccentricity ($0.023$), near synchronisation ($P_{\rm bin}$= \SI{7.448}{\day}
and $P_{\rm rot}$= \SI{7.775}{\day}) and spin--orbit alignment ($<20^{\circ}$).

\subsection{Differential rotation and true obliquity}
Although unlikely, it is
    worth mentioning that the true stellar obliquity may be non-zero, i.e. that
    the star is pointing either towards or away from us and is thus still
    realigning. Ordinarily, an independent measurement of the rotation period
    from spot modulation, together with a projected equatorial rotational
    velocity from the Rossiter-McLaughlin analysis, can provide a measurement of
    the stellar inclination. However, \citet{kostov2020} show that neither
    \tess{} nor ASAS-SN photometry display any clear periodicity. We show our
    fit to the differential rotation model in the right panel of
    Fig.~\ref{fig:fit}. The fit favours a higher stellar
inclination, $\istar{} = 137.1^{+13.7}_{-74.8}\,\si{\degree}$, 
with true obliquity $\psi = \ang[angle-symbol-over-decimal]{44.8} \pm
\ang[angle-symbol-over-decimal]{8.8}$ and latitudes rotating antisolar with
$\alpha < 0$ at the 99th percentile. The implied rotation period here is
$17.3^{+3.7}_{-3.2}\,$\si{\day}, 
which is similar to the rigid body case. The fit improves the Bayesian
Information Criterion by \num{5.5}, showing an improvement in the model,
although the reduced $\chi_\nu^2 = 0.45$ suggests we may be over-fitting the
data. There have only been detections of antisolar differential rotation in
three giant stars \citep{strassmeier2003,kovari2013,kriskovics2014}, with only a
small handful of main sequence Sun-like candidates \citep{benomar2018}.
Ultimately, we are not able to rule out contributions from correlated noise and
therefore consider it unlikely that the star is inclined. We conclude that the
rigid body model is the most likely scenario and thus the true obliquity of the
star is low.

\subsection{Spectral contamination from secondary star}

\begin{figure}
    \centering
    \includegraphics{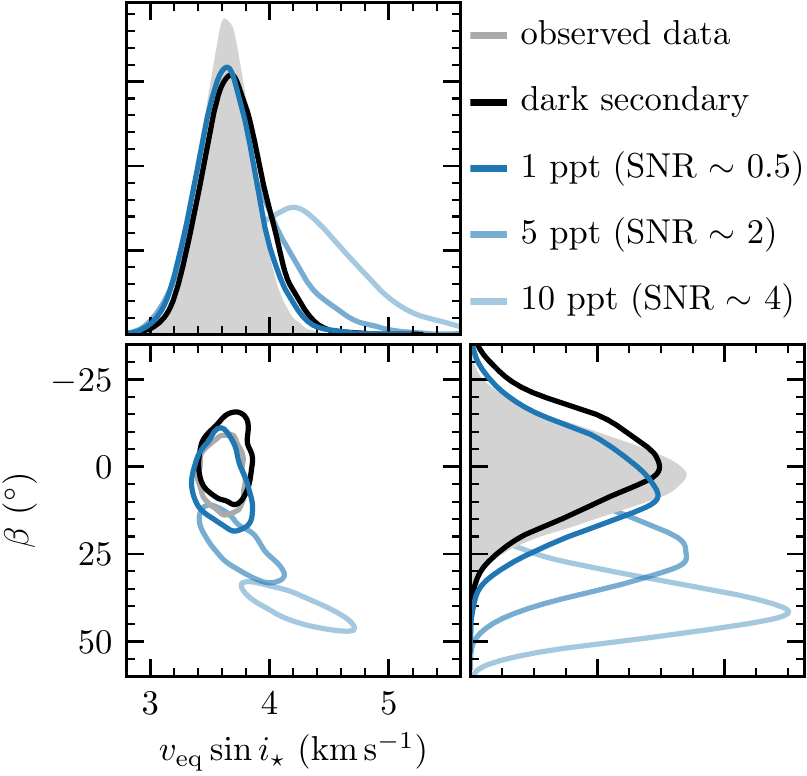}
    \caption{Simulation of the impact of varying levels of emission flux from
    the secondary star on the retrieved values for \vsini{} and $\beta$. The
posterior distribution for the observed data is shown in grey, and is compared
to the simulation with no contamination from a secondary spectrum in black. The
coloured lines show the simulated effect for increasing relative flux
contribution from the occulting body. The estimated flux contamination for
J0608-59 is  \SI{\sim 1}{\ppt}, which agrees very well with the simulated data.
The lower left panel shows the $1\sigma$ contours.}
    \label{fig:simulation}
\end{figure}

The systems that make up the EBLM sample (including J0608-59) are single-lined
binaries (SB1), where the spectrum from the secondary star is too faint to be
detected in individual observations. However, if the flux emission from the
secondary star exceeds that of the noise in the data, the faint signal of the
secondary spectrum may be imprinted in the spectrum that is dominated by the
light from the primary star, and impact the shape of the final CCF. Since the
reloaded Rossiter-McLaughlin method depends on detecting changes in the CCF
shape during eclipse, this may impact our measurements. It is worth noting that
for the vast majority of the binary orbit this is still not a problem, as the
relative radial velocity motion between the two stars is so large that the line
core from the secondary star falls outside the line core of the primary star and
will not affect the CCF shape. Most of the contamination will occur close to
conjunctions, where their radial velocity curves cross.

The apparent magnitude of the system in the \coralie{} $V$ band is $m_{V,1} =
11.73$. Assuming an age of \SI{5}{\gyr}, \citet{triaud2017} predict an apparent
brightness for the secondary star $m_{V,2} = 19.75$, which gives an estimated
secondary-to-primary flux ratio of \SI{\sim 1}{\ppt} through the relation
\begin{align*} \Delta m_V = -2.5 \log{\left(\frac{F_{V,2}}{F_{V,1}}\right)}.
\end{align*} The fitted noise of the residual CCFs in Fig.~\ref{fig:trace} is
\SI{\sim 2.5}{\ppt}, which yields a SNR \SI{<0.5}{} for the spectrum from the
secondary star. The impact of the secondary spectrum should therefore not affect
our measurements of \vsini{} and $\beta$. Despite this, we carry out a
simulation to quantify the impact of the secondary spectrum, and how it changes
with increasing flux ratio. We simulate a series of CCFs for the primary star
during eclipse, centered at the radial velocity of the primary star in the
barycentric reference frame, assuming Gaussian profiles with depths and widths
determined by their typical observed values. We add uncertainties to the CCFs
according to the typically fitted error, $\epsilon$ from the MCMC analysis in
Section~\ref{sec:methods_occulted_light}. We add a CCF from the secondary star,
where we assume that the contrast and FWHM is the same as for the primary, with
flux ratio $\delta F$, and centered at the predicted radial velocity of the
secondary star from the Keplerian orbit, again in the barycentric reference
frame. Finally, we assume Gaussian profiles for the distortion due to the
Rossiter-McLaughlin effect, with contrasts and FWHM as fitted in
Section~\ref{sec:methods_occulted_light}, and their radial velocity centres
translated to the barycentric reference frame. This contribution is subtracted
from our combined CCF from the two stars to simulate the missing light from the
occulted disc.

We apply the reloaded Rossiter-McLaughlin effect outlined in
Section~\ref{sec:analysis} to our simulated CCFs, which, as described above,
include contributions from both stars and the distortion of the line profiles
due to an occulted disc. We vary the flux ratio of the secondary using $F_2/F_1
= \{1, 5, 10\}\,\si{\ppt}$, and repeat the MCMC sampling procedure from before
to obtain posterior distributions for \vsini{} and $\beta$. The posterior
distributions are shown in Fig.~\ref{fig:simulation}. The simulation with
$F_2/F_1 = \SI{1}{\ppt}$ matches the observed data remarkably well, and does not
show any significant bias in the derived parameters when compared to the
simulation with a dark secondary. However, the posterior distributions for both
parameters become significantly biased for higher flux ratios. Our simulations
show that care must be taken for eclipse spectroscopy with luminous secondaries,
and potentially even for eclipse spectroscopy of exoplanets due to the hot
day-side temperatures of some ultra-hot Jupiters, which can reach similar
temperatures to M or even K dwarfs. Assuming a flux ratio of \SI{\sim 15}{\ppt},
as for the Kepler-16 system \citep{doyle2011a}, the secondary spectrum could
bias the measurement of $\beta$ to   \SI{>30}{\degree} at $3\sigma$
significance, and overestimate \vsini{} by $\simeq$\SI{30}{\percent}

A few caveats are worth mentioning with the simulations. First, the relative
depth and width of the secondary CCF are assumed to be the same as for the
primary CCF. This may not always be the case when the spectral types of the
stars are very different. Taking a G-M binary as an example, the convolution of
an M dwarf spectrum with a G2 mask can yield shallower (thus lower SNR) CCFs
than assumed here due the cross-correlation with template lines that are not
present in M dwarf spectra. In practice the simulations we carry out here
presents a worst-case scenario. Second, the impact of the secondary spectrum
will also depend on the specific orbital characteristics, in particular the
eccentricity and argument of periastron, which determine where and how much the
two spectra will overlap.

\section{Conclusion} We have presented a stellar obliquity measurement of the
primary star in the EBLM J0608-59/TOI-1338 eclipsing binary system, recently
discovered to host a circumbinary planet. High-resolution spectroscopy during
primary eclipse supports a low obliquity, consistent with alignment with the
binary and planet orbits. The binary orbit has not pseudo-synchronised, which
indicates that the obliquity of the star has not been influenced by tides, and
is thus likely primordial. Moreover, we have simulated the effect of an
unresolved secondary spectrum on our analysis of eclipse spectra, and show that
the effect is negligible for our data.  However, secondaries that contribute
with flux of just a few thousandths of that of their primaries can strongly bias
measurements of \vsini{} and $\beta$.

J0608-59 is but one of \num{\sim20} EBLM systems for which we are currently
studying stellar obliquities from Rossiter-McLaughlin sequences observed with
\coralie{} and \harps{}, and currently the only one known to host a circumbinary
planet.  However, our sample has significant overlap with the BEBOP Doppler
survey for circumbinary planets, therefore we may soon find more similar
systems. In a forthcoming paper we will present obliquity measurements on
\num{\sim20} systems observed with \coralie{} and \harps{}, which will allow us
to place constraints on tidal evolution in low-mass binaries and their
circumbinary planets.

\section*{Acknowledgements}

We are very grateful to the anonymous referee for carefully reading our
manuscript and providing constructive comments which led to additional analysis
that substantially improved the paper.  This work includes data obtained from
\coralie{}, an instrument mounted on the Swiss \euler{} 1.2m telescope, a
project of the University of Geneva, funded by the Swiss National Science
Foundation.
This work was in part funded by the U.S.--Norway Fulbright Foundation and a NASA
\tess{} GI grant G022253 (PI: Martin). AHMJT  received funding from the European
Research Council (ERC) under the European Union’s Horizon 2020 research and
innovation programme (grant agreement n$^\circ$ 803193/BEBOP), and from a
Leverhulme Trust Research Project Grant (RPG-2018-418). VKH is also supported by
a Birmingham Doctoral Scholarship, and by a studentship from Birmingham's School
of Physics \& Astronomy. DVM received funding from the Swiss National Science
Foundation (grant number  P 400P2 186735). SG has been supported by STFC through
consolidated grants ST/L000733/1 and ST/P000495/1.

\section*{Data availability} The data underlying this article will be shared on
reasonable request to the corresponding author.

\bibliographystyle{mnras}
\bibliography{references}




\bsp	
\label{lastpage}
\end{document}